\documentclass[12pt,a4paper]{article}
\usepackage[latin1]{inputenc}
\usepackage{amsmath}
\usepackage{amsfonts}
\usepackage{amssymb}
\usepackage{hyperref}
\hypersetup{colorlinks=true}
\usepackage{cite}
\begin{document}
\author{Levent Akant${^\dag}$, Emine Ertu\u{g}rul${^\ddag}$,\\ Yusuf G\"{u}l${^\S}$, O. Teoman Turgut${^*}$ \\ Department of Physics, Bo\u{g}azi\c{c}i University \\ 34342 Bebek, Istanbul, Turkey \\ $^\dag$levent.akant@boun.edu.tr, $^\ddag$emine.ertugrul@boun.edu.tr,
\\${^\S}$yusufgul.josephrose@gmail.com,${^*}$turgutte@boun.edu.tr}
\title{\bf Boundary Effects on Bose-Einstein Condensation in Ultra-Static Space-Times}
\maketitle
\begin{abstract}
The boundary effects on the Bose-Einstein condensation with a nonvanishing chemical potential on an ultra-static space-time are studied. High temperature regime, which is the relevant regime for the relativistic gas, is studied through the heat kernel expansion for both Dirichlet and Neumann boundary conditions. The high temperature expansion in the presence of a chemical potential is generated via the Mellin transform method as applied to the harmonic sums representing the free energy and the depletion coefficient. The effects of boundary conditions on the relation between the depletion coefficient and the temperature are analyzed. Both charged and neutral bosons are considered.
\end{abstract}

\newpage

\section{Introduction}
In this paper we will derive finite size effects on the Bose-Einstein condensation of a Bose gas on an ultra-static space time $\mathbf{R}\times \mathbf{M}$ in the limit $\beta m \rightarrow 0$. The Bose gas will be confined in a box $B$ of arbitrary shape with volume $V$ and surface area $A$, and it will be subject to either Dirichlet or Neumann boundary conditions. We will consider both charged and neutral gases. The metric on $\mathbf{R}\times \mathbf{M}$ is
\begin{equation}
  ds^{2}=-dt^{2}+\gamma_{ij}dx^{i}dx^{j}.
\end{equation}
Clearly the Minkowski space is a special case.

   The first step in our analysis will be to express the free energy $A$ and the depletion coefficient $N_{e}$ for the condensate in the presence of a non-vanishing chemical potential $\mu\neq 0$ as harmonic sums i.e. as sums of the form $\sum_{k} f(k\beta m)$, involving the trace of the heat kernel. It is well known that in the relativistic regime it is the small values of $\beta m$ which is relevant for the Bose-Einstein condensation. The natural way to develop the asymptotic expansion of a harmonic sum is to take the Mellin transform, determine the poles and residues of an appropriate meromorphic extension of the transform, and finally take the inverse transform (Mellin-Barnes integral) of the singular terms \cite{Ober, Wong, Flajolet}. The next step will be the derivation of the small $\beta m$ asymptotic expansion (high temperature expansion) of $A$ and $N_{e}$  by applying the Mellin transform techniques to the harmonic sums representing $A$ and $N_{e}$. No use will made of zeta function regularization which is the basis of the conventional approach to the high temperature expansion \cite{Dowker1,Actor,Dowker2,Dowker3,Kirsten1,Kirsten2,T1,T2}. Thus our analysis in the present work will provide an alternative way of regularizing and developing the high temperature expansion.  Next, we will use this high temperature expansion to study the effects of boundaries on the Bose-Einstein condensation on a manifold. It will be shown that in three dimensions and in the charged case the leading finite size correction is logarithmic in $\beta m$; whereas in the neutral case it will be a power law. Higher dimensional case will also be discussed. Our results for the free energy will be in accord with the ones derived in \cite{Kirsten1,T1,T2}. The usual objection against the use of Mellin transform in the case $\mu\neq 0$ (see \textit{e.g.} Sec. 5.4 of \cite{tomsbook} ) is a potential divergence caused by exponentially increasing factor of $e^{\mu \beta}$ appearing in the transform. However we will show that large time decay properties of the heat kernel \cite{LiBook} imply that in fact the Mellin transform remains finite and the high temperature expansion proceeds without any divergence problem.

Special relativistic Bose-Einstein condensation in thermodynamic limit was discussed by many authors \cite{Jutner, Glaser, Landsberg, Nieto, Altaie,
Beckmann, Carvalho, Beckmann2, Haber, Haber2, Burakovsky, Filippi}, the interacting case in flat space was studied in \cite{Dodelson}, and the curved space generalization was given in \cite{TK}. In curved space, the relation between condensation and symmetry breaking was discussed in \cite{T1,T2,T3,TS}. The finite size effects in the non-relativistic flat case were studied in \cite{Pathria1, Pathria2, Pathria3, Pathria4, Pathria5, Pathria6, Paj}, those in the special relativistic case in \cite{Pathria7, Singh, Hu}, and those in the non-relativistic curved case in \cite{T4}. Finite size effects in relativistic gas of neutral bosons were presented in \cite{Begun0, Begun}. A discussion of the
Bose gas inside a harmonic trap through Mellin transform can be found in \cite{Kirsten3}. For an approach to the Bose-Einstein condensation for both noninteracting and interacting systems on a manifold based on the global heat kernels estimates of \cite{LiYau} see \cite{Biz}.

Here is the brief summary of the present paper. In Sec. 2 we shall discuss the relation between the asymptotic expansions of harmonic sums and Mellin transforms and shall work out the high temperature expansion of the free energy for non-vanishing chemical potential using this method. In Sec. 3 we shall analyze the net charge of excited bosons using similar techniques and derive the finite size effects on the relation between temperature and the charge density. In Sec. 4 we shall consider the free energy of the neutral Bose gas along the same lines, and in Sec 5. we shall investigate the effects of boundaries on the relation between temperature and number density of excited neutral bosons. In Appendix A we shall derive the asymptotic expansion of a certain function that is of fundamental importance in our analysis. In Appendix B we shall discuss further technicalities involved in the use of the Mellin transform to generate the high temperature asymptotic expansion.

\section{Free Energy of Charged Bosons}

The Klein-Gordon equation (KG) on $\mathbf{R}\times \mathbf{M}$ with external potential $V$ is
\begin{equation}
  (-\Box+m^{2}+U)\psi=0,
\end{equation}
where
\begin{equation}
  \Box=\frac{1}{\sqrt{g}}\partial_{\mu}\sqrt{g}g^{\mu\nu}\partial_{\nu}.
\end{equation}
Separating the time derivative we can write KG as
\begin{equation}
  -\frac{\partial^{2}\psi}{\partial t^{2}}=\left[-\Delta+U+m^{2}\right]\psi,
\end{equation}
where
\begin{equation}
  \Delta=\frac{1}{\sqrt{\gamma}}\partial_{i}\sqrt{\gamma}\gamma^{ij}\partial_{j}
\end{equation}
is the Laplacian on $\mathbf{M}$.
We will define the single particle Hamiltonian as
\begin{equation}
  H=-\Delta+U+m^{2}.
\end{equation}
We will assume that $-\Delta+U$ is a non-negative operator. This clearly implies $H$ is positive. If the former operator fails to be non-negative we will still assume that $H$ is positive. In the latter case without loss of generality we can redefine $U$ and $m^{2}$ in such a way that the new $-\Delta+U$ is non-negative and the new $m^{2}$ is positive.

Free Energy of the charged boson is;
\begin{equation}\label{fchr}
A=\frac{1}{\beta} \sum \limits_{\sigma} \left[ \log \left( 1- e^{-(\beta^{2} \lambda_{\sigma}+\beta^{2}m^{2})^{1/2} + \beta\mu} \right) + \log \left( 1- e^{-(\beta^{2} \lambda_{\sigma}+\beta^{2}m^{2})^{1/2} - \beta\mu} \right)  \right].
\end{equation}
Here $\lambda_{\sigma}$'s are the eigenvalues of $-\Delta+U$ which by assumption are non-negative.

After using
\begin{equation}
- \log(1-x)= \sum \limits_{k=1} ^{\infty} \frac{x^{k}}{k}
\end{equation}
and the subordination identity
\begin{eqnarray}\label{sub}
  e^{-b\sqrt{x}} &=&
  \frac{b}{2\sqrt{\pi}}\int_{0}^{\infty}\frac{dt}{t^{3/2}}\,e^{-\frac{b^{2}}{4t}}\,e^{-tx},
\end{eqnarray}
in (\ref{fchr}), and changing the integration variable to $t=s\beta^{2}m^{2}$ we get the free energy of the system as;
\begin{eqnarray}\label{A}
A = -\sum\limits_{k=1}^\infty \frac{m  }{\sqrt{\pi}}\cosh(k\beta \mu)\,F(k\beta m)
\end{eqnarray}
where
\begin{equation}\label{F}
 F(x)=\int_{0}^{\infty}\frac{dt}{t^{3/2}}\, e^{\frac{-x^{2}}{4t}}e^{-t}\,(Tr\, e^{\frac{t}{m^{2}}(\Delta-U)}).
\end{equation}
The right hand side of (\ref{A}) is a series of the form
\begin{equation}\label{hs}
A(x)= -\frac{m  }{\sqrt{\pi}}\sum_{k=1}^{\infty}g(kx),
\end{equation}
where
\begin{equation}
  g(x)=\cosh (\xi x)\,F(x),\;\;\;x=\beta m,\;\;\;\xi=\frac{\mu}{m}.
\end{equation}
Such series are known as harmonic sums and their $x\rightarrow 0$ asymptotics can be determined from their Mellin transforms. The Mellin transform of a function $f(x)$ is defined as
\begin{equation}
   \widetilde{f}(s)=(\mathcal{M}f)(s)=\int_{0}^{\infty}dx\,x^{s-1}f(x).
\end{equation}
A straightforward calculation shows that
\begin{equation}
  (\mathcal{M}A)(s)=-\frac{m  }{\sqrt{\pi}}\zeta(s)(\mathcal{M}g)(s).
\end{equation}
Here $\zeta(s)$ is the Riemann zeta function which is a meromorphic function on the whole complex plane with a single simple pole at $s=1$.

In general the Mellin transform $(\mathcal{M}f)(s)$ of a function $f$ is a holomorphic function of $s$ in a vertical strip $\langle a,b\rangle$ bounded by the vertical lines $s=a$ and $s=b$. Largest such strip is called the fundamental strip of the Mellin transform. The small $x$ asymptotic of $f(x)$ is determined by the poles and residues of a meromorphic extension (if it exists) of the Mellin transform beyond the fundamental strip. This is done via the inverse Mellin (Mellin-Barnes) transform,
\begin{equation}
  (\mathcal{M}^{-1}\widetilde{f})(x)=\int_{c-i\infty}^{c+i\infty}ds\,\widetilde{f}(s)x^{s},
\end{equation}
where the vertical line $s=c$ lies in the fundamental strip. One has to complete this contour into a closed rectangular one enclosing the poles of the meromorphic extension of $(\mathcal{M}f)(s)$, then performing the integral one gets the asymptotic expansion of $f(x)$. The details of the method can be found in \cite{Ober, Wong, Flajolet}. The correspondence between the poles of $(\mathcal{M}f)(s)$ and the asymptotics of $f(x)$ is as follows. If the Mellin transform has the singular expansion with residues $A(w,k)$
\begin{equation}
  (\mathcal{M}f)(s)\asymp\sum_{w,k}\frac{A(w,k)}{(s-w)^{k+1}},
\end{equation}
where $\asymp$ refers to the singular part of a function, then
\begin{equation}
 f(s)\sim \sum_{w,k}A(w,k)\frac{(-1)^{k}}{k!}t^{-w}(\log t)^{k},
\end{equation}

Notice the presence of logarithmic terms which arise from multiple poles. In order for this to be really an asymptotic expansion one must make sure that the contributions coming from the extra horizontal and vertical lines used to close the contour are negligible. There are several different sufficient conditions which guarantee this. A sufficient condition suitable for our purposes will be
\begin{equation}
  \int_{-\infty}^{\infty} d \tau\, |\widetilde{f}(\sigma+i\tau)|<\infty.
\end{equation}
Here $\sigma$ is a real number lying between two consecutive poles of $\widetilde{f}(s)$ with distinct real parts. This part of the problem will be discussed in the Appendix.

Before we apply this general method to (\ref{N}) we must determine the $x\rightarrow 0$ singular asymptotic of $F(x)$, as will be shown below this will be crucial in the determination of the poles of $(\mathcal{M}A)(s)$. The derivation of this singular asymptotics will be given in the Appendix A, here we just quote the result:
\begin{equation}\label{singF}
  F(x)\asymp\sum_{n=1}^{4}b_{-n}x^{-n}+b'\log (x/2).
\end{equation}
Here the coefficients $b$ are given in terms of the heat kernel coefficients of $e^{t\Delta}$.
If
\begin{equation}
  Tr e^{t\Delta}\sim \frac{1}{t^{3/2}}\sum_{k=0}^{\infty} a_{k/2}t^{k/2}
\end{equation}
then
\begin{eqnarray}\label{b}
  b_{-4} &=& 16 m^{3} a_{0} \\
  b_{-3} &=& 4\sqrt{\pi}m^{2}a_{1/2} \\
  b_{-2} &=& 4(-m^{3}a_{0}+ma_{1}) \\
  b_{-1} &=& 2(-\sqrt{\pi}m^{2}a_{1/2}+\sqrt{\pi}\,a_{3/2})\\
  b' &=& -m^{3}a_{0}+2ma_{1}-m^{-1}a_{2}.
\end{eqnarray}
The behavior of $F(x)$ for large $x$ will also be needed in the following discussion. Let $\lambda$ denote the first nonzero eigenvalue of $-\Delta$ subject to either Dirichlet of Neumann boundary conditions. If the smallest eigenvalue of $H$ is nonzero then according to the Lemma 14.1 of \cite{LiBook} we have
\begin{equation}\label{in}
    Tr e^{-H\,t} \leq
    Tr (e^{-H\,t_{0}})\,e^{-\lambda(t-t_{0})},\;\;\;\textrm{for}\;\;\;\; t\geq t_{0}.
    \end{equation}
Using the heat kernel asymptotics
\begin{equation}
  Tr e^{-H\,t_{0}}\sim \frac{V}{(4\pi t_{0})^{3/2}},\;\;\;\;t_{0}\rightarrow 0
\end{equation}
we see that by choosing $t_{0}$ small enough we get
\begin{equation}\label{trineq}
      Tr e^{-H\,t} \leq \frac{3}{2}\frac{V}{(4\pi _{0}t)^{3/2}}.
\end{equation}
Using this bound in (\ref{F}) we obtain the bound
\begin{equation}\label{ineq1}
  F(x)\leq \frac{V}{(4\pi t_{0})^{3/2}}\frac{3\sqrt{2}}{x}K_{1/2}(x)=\frac{V}{(4\pi t_{0})^{3/2}}\frac{3\sqrt{2}}{x}{\sqrt{\frac{\pi}{2x}}}e^{-x}.
\end{equation}
Here $K_{1/2}(x)$ is the modified Bessel function of order $1/2$. Here we used
\begin{equation}
  K_{1/2}(x)={\sqrt{\frac{\pi}{2x}}}e^{-x}.
\end{equation}
Thus
\begin{equation}\label{largeF}
  F(x)=O\left(x^{-3/2}e^{-x}\right),\;\;\;\;x\rightarrow\infty.
\end{equation}

If the smallest eigenvalue of $H$ is vanishing, as in the case of Neumann problem for the Laplacian, we again have from Lemma 14.1 of \cite{LiBook}:
\begin{equation}
  Tr e^{-H\,t}\leq 1+\left(Tr\,e^{-H\,t_{0}}-1\right)\,e^{-\lambda(t-t_{0})}.
\end{equation}
Proceeding as above we again get (\ref{ineq1}) and (\ref{largeF}).



We will see shortly that $(\mathcal{M}g)(s)$ is holomorphic for $\Re s > 4$. Now  the main problem is the presence of exponentially increasing $\cosh$ term which makes it difficult to meromorphically continue $(\mathcal{M}g)(s)$ to $\Re s < 4$. In order to find the meromorphic extension we will write $A$ as
\begin{equation}
 A=-\frac{m}{\sqrt{\pi}} \sum\limits_{k=1}^\infty e^{-\alpha k x}g_{\alpha}(kx)
\end{equation}
where
\begin{equation}
  g_{\alpha}(x)= e^{\alpha  x}  \cosh(\xi x) F(x).
\end{equation}
Here $\alpha$ is an arbitrary parameter with $0<\alpha<1-\xi$, which will be used as a regularizing parameter in the integrals arising in the derivation of the asymptotic expansion.

Using (\ref{singF}) we see that as $x\rightarrow 0$
\begin{equation}\label{asy}
  g_{\alpha}(x)\sim d_{-4}x^{-4}+c_{-3}x^{-3}+d_{-2}x^{-2}+d_{-1}x^{-1}+d_{0}+\ldots
\end{equation}
and using (\ref{largeF}) we see that as $x\rightarrow \infty$
\begin{equation}\label{largeg}
  g_{\alpha}(x)=O\left(x^{-3/2}\,e^{-(1-\xi-\alpha)x}\right),
\end{equation}
where the coefficients $d$ are related to the coefficients $b$. The first few are
\begin{eqnarray}\label{d}
  d_{-4} &=&  b_{-4} \\
  d_{-3} &=&  b_{-3}+\alpha b_{-4} \\
  d_{-2} &=&  b_{-2}+ \alpha b_{-3}+ \frac{(\alpha^{2}+\xi^{2})}{2} b_{-4}  \\
  d_{-1} &=&  b_{-1}+ \alpha b_{-2}+ \frac{(\alpha^{2}+\xi^{2})}{2} b_{-3}+\frac{(\alpha^{3}+3\alpha \xi^{2})}{6} b_{-4}\\
  d_{0} &=&  \alpha b_{-1}+ \frac{(\alpha^{2}+\xi^{2})}{2} b_{-2}+ \frac{(\alpha^{3}+3\alpha \xi^{2})}{6}b_{-3} +\frac{(\alpha^{4}+6\alpha^{2}\xi^{2}+\xi^{4})}{24} b_{-4} \nonumber\\
  \end{eqnarray}
Let's also recall the first few heat kernel coefficients \cite{Gilkey}
\begin{eqnarray}
  a_{0} &=& \frac{V}{(4\pi)^{3/2}} \\
  a_{1/2} &=& r \frac{A}{16\pi} \\
  a_{1} &=&  \frac{1}{6(4\pi)^{3/2}}\left[\int_{B}d\mu_{g}(-6U-R)+\int_{\partial B}2K\right]
\end{eqnarray}
Here $R$ is the Riemann curvature of $\mathbf{M}$, $K$ is the extrinsic curvature of the surface $\partial B$ of the box, and $d\mu_{g}$ is the Riemann volume element on $\mathbf{M}$.
 In $a_{1/2}$, $r=1,-1$ for Neumann and Dirichlet boundary conditions, respectively.
Now the Mellin transform of $A$ is given by
\begin{equation}
  (\mathcal{M} A)(s)=\frac{-m}{\sqrt{\pi}}\zeta(s)(\mathcal{LM}g_{\alpha})(s,\alpha).
\end{equation}
Here $\mathcal{LM}$ is the Laplace-Mellin transform \cite{JorLang}
\begin{equation}\label{LM}
  (\mathcal{LM}g_{\alpha})(s,\alpha)=\int_{0}^{\infty}dx\,x^{s-1}\,e^{-\alpha x}g_{\alpha}(x).
\end{equation}
From the behavior of the integrand for large $x$ (\ref{largeg}) we conclude that the upper limit of integration is not problematic. On the other hand for $\alpha \geq 0$, $g(x)=O(x^{-4})$ as $x\rightarrow 0$ and we see that the integral is convergent for $\Re s>4$.  Write $ (\mathcal{LM}g_{\alpha})(s,\alpha)$ as
\begin{eqnarray}
  (\mathcal{LM}g_{\alpha})(s,\alpha)&=&\int_{0}^{\infty}dx\,x^{s-1}\,e^{-\alpha x}\left[ g_{\alpha}(x)-d_{-4}x^{-4}\right]-\int_{0}^{\infty}dx\,x^{s-1}\,e^{-\alpha x}d_{-4}x^{-4}\nonumber\\
    &=&\int_{0}^{\infty}dx\,x^{s-1}\,e^{-\alpha x}\left[ g_{\alpha}(x)-d_{-4}x^{-4}\right]+d_{-4}\alpha^{4-s}\Gamma(s-4).
\end{eqnarray}
 From (\ref{asy}) we get $ g_{\alpha}(x)-d_{-4}x^{-4}=O(x^{-3})$ and we see that the lower limit of the integral is now finite for $\Re s>3$. On the other hand from (\ref{largeg}) we see that the upper limit of the integral is also finite. From the second integral in the above expression we get a pole at $s=4$. Thus we get a meromorphic continuation of $(\mathcal{LM}g)(s,\alpha)$ to the region $\Re s>3$. Continuing in this manner and subtracting more and more terms of the asymptotic series (\ref{asy}) from $g(x)$ we get a meromorphic extension of $(\mathcal{LM}g)(s,\alpha)$ into the half-plane to the left of $\Re s=4$. The resulting expansion is
\begin{eqnarray}
  (\mathcal{LM}g_{\alpha})(s,\alpha)\sim d_{-4}\alpha^{4-s}\Gamma(s-4)+d_{-3}\alpha^{3-s}\Gamma(s-3)+d_{-2}\alpha^{2-s}\Gamma(s-2) \nonumber\\
  + d_{-1}\alpha^{1-s}\Gamma(s-1)+ d_{0}\alpha^{-s}\Gamma(s)\ldots
\end{eqnarray}
Note that all the poles are simple. The pole at $s=4$ comes from the $d_{-4}$ term, while the one at $s=4$ comes from both $d_{-4}$ and $d_{-3}$ terms, etc.

Coming back to $(\mathcal{M}A)(s)=-m \pi^{-1/2}\zeta(s)(\mathcal{LM}g)(s,\alpha)$ we see that the poles are the same as above except that the pole at $s=1$ is now a double pole because of the Riemann $\zeta$-function which behaves as
\begin{equation}\label{zeta}
\zeta(s)\sim \frac{1}{s-1}+\gamma,\;\;\;s\rightarrow 1.
\end{equation}
Here $\gamma$ is the Euler-Mascheroni constant. On the other hand we also have the singular expansion
\begin{equation}
  \Gamma(s-n)\asymp \sum_{m=0}^{\infty} \frac{(-1)^{m}}{m!}\frac{1}{s-n+m}.
\end{equation}
Using these we get the behaviour of $(\mathcal{M}A)(s)$ around its poles and the corresponding asymptotic expansion of $A$.

Let us write
\begin{equation}
  A\sim A^{(4)}+A^{(3)}+\ldots
\end{equation}
where $A^{(m)}$ is the contribution from the pole at $s=m$.

For $s=4$
\begin{equation}
  (\mathcal{M}A)(s)\asymp \frac{-m}{\sqrt{\pi}}\zeta(4)b_{-4}\,\frac{1}{s-4}
\end{equation}
Thus the corresponding asymptotic is
\begin{eqnarray}
  A^{(4)}= \frac{-m}{\sqrt{\pi}} \zeta(4)b_{-4}(\beta m)^{-4}
\end{eqnarray}

For $s=3$
\begin{equation}
  (\mathcal{M}A)(s)\asymp \frac{-m}{\sqrt{\pi}} \frac{\zeta(3)}{s-4}\, b_{-3}
\end{equation}
Thus the corresponding asymptotic is
\begin{eqnarray}
  A^{(3)}= \frac{-m}{\sqrt{\pi}} \zeta(3) b_{-3} (\beta m)^{-3}
\end{eqnarray}

For $s=2$
\begin{equation}
  (\mathcal{M}A)(s)\asymp \frac{-m}{\sqrt{\pi}} \frac{\zeta(2)}{s-2}\, \ \left(\frac{\xi^{2}}{2}b_{-4}+b_{-2}\right)
\end{equation}
Thus the corresponding asymptotic is
\begin{eqnarray}
  A^{(2)}= \frac{-m}{\sqrt{\pi}} \zeta(2) \left(\frac{\xi^{2}}{2}b_{-4}+b_{-2}\right) (\beta m)^{-2}
\end{eqnarray}

For $s=1$
\begin{eqnarray}
(\mathcal{M}A)(s) &\sim & \frac{-m}{\sqrt{\pi}}\left(\frac{1}{s-1}+\gamma\right)\left(FP(\mathcal{LM}g)(s)|_{s\rightarrow 1}+\frac{\left[ \frac{\xi^{2}}{2} b_{-3}+ b_{-1}\right]}{s-1}   \right) \nonumber\\
 &\asymp & \frac{-m}{\sqrt{\pi}} \left[ \frac{\left[ \frac{\xi^{2}}{2} b_{-3}+b_{-1}\right]}{(s-1)^{2}} \right. \nonumber\\
 & + & \left. \frac{FP(\mathcal{LM}g)(s=1,\alpha)+ \gamma \left[ \frac{\xi^{2}}{2} b_{-3}+ b_{-1}\right] }{s-1}\right].
\end{eqnarray}
Here $FP$ means the finite part of the expression on the right
\begin{eqnarray}
  FP(\mathcal{LM}g)(s=1,\alpha)&=&(\mathcal{LM}g)(s=1,\alpha)-\frac{\left[ \frac{\xi^{2}}{2} b_{-3}+ b_{-1}\right]}{s-1}\nonumber\\
  &=&\int_{0}^{\infty}dx\,x^{s-1}\cosh(\xi x)F(x)- \frac{\left[ \frac{\xi^{2}}{2} b_{-3}+ b_{-1}\right]}{s-1}.
\end{eqnarray}
So
\begin{eqnarray}
  A^{(1)} &=& \frac{m}{\sqrt{\pi}}\left[ \frac{\xi^{2}}{2} b_{-3}+ b_{-1}\right](\beta m)^{-1}\log (\beta m)^{-1} \nonumber\\
  &&+\frac{-m}{\sqrt{\pi}} \left( \gamma \left[ \frac{\xi^{2}}{2} b_{-3}+ b_{-1}\right]+FP(\mathcal{LM}g)(1,\alpha) \right)  (\beta m)^{-1}. \nonumber\\
 \end{eqnarray}


Finally expressing all the $b$ coefficients in terms of heat kernel coefficients through (\ref{d}) and (\ref{b}) we get:
\begin{eqnarray}
 A^{(4)} &=& \frac{-16}{\sqrt{\pi}} \zeta(4) m^{4} a_{0} (\beta m)^{-4} \nonumber\\
 A^{(3)} &=& -4 \zeta(3) m^{3} a_{1/2} (\beta m)^{-3} \nonumber\\
 A^{(2)} &=& -\frac{4m^{2}}{\sqrt{\pi}}\zeta(2) \left[ (2\mu^{2}-m^{2})a_{0}+a_{1}\right] (\beta m)^{-2} \nonumber\\
 A^{(1)} &=& \left[ 2 (\mu^{2} m-m^{3}) a_{1/2}+ 2 m\, a_{3/2})\right]  (\beta m)^{-1} \log (\beta m)^{-1} \nonumber\\
 &-&  \left( \gamma \left[  2(\mu^{2}m-m^{3}) a_{1/2}+2 m\, a_{3/2} \right]   +\frac{m}{\sqrt{\pi}}FP(\mathcal{LM}g)(1,\alpha) \right)  (\beta m)^{-1}. \nonumber\\
\end{eqnarray}
Note that the result does not depend on the regularization parameter $\alpha$. This result is in accord with \cite{Kirsten1,T1,T2}.

The above analysis can be repeated in dimension $d>3$ with the result
\begin{eqnarray}
A &\sim &  -\zeta(d+1) \,\frac{(2m)^{d+1}}{\sqrt{\pi}}  \, \Gamma\left(\frac{d+1}{2} \right) \,a_{0}\,(\beta m)^{-(d+1)} \nonumber \\
   &-& \zeta(d)\, \frac{(2m)^{d}}{\sqrt{\pi}} \,\Gamma\left(\frac{d}{2} \right) \, a_{1/2} \,(\beta m)^{-d} \nonumber \\
   &-& \zeta(d-1) \,\frac{(2m)^{d-1}}{\sqrt{\pi}}\,\Gamma\left(\frac{d-1}{2} \right)\, \left(\left[-m^{2}\, +(d-1)\,\mu^{2}\right] a_{0} + a_{1} \right)\, (\beta m)^{-(d-1)}.\nonumber\\
\end{eqnarray}

\section{Charge}
Using the subordination identity (\ref{sub}) the net charge of the excited particles can be written as
\begin{eqnarray}\label{N}
 Q_{e}&=&\sum\limits_{\sigma\neq 0}\left[\frac{1}{e^{\beta( \lambda_{\sigma}+m^{2})^{1/2}-\beta\mu}-1}-
    \frac{1}{e^{\beta(
    \lambda_{\sigma}+m^{2})^{1/2}+\beta\mu}-1}\right]\\
& =& \sum\limits_{k=1}^\infty \frac{(k\beta m)\sinh(k\beta \mu)}{\sqrt{\pi}}F_{1}(k\beta m).
\end{eqnarray}
Here
\begin{equation}\label{F1}
 F_{1}(x)=\int_{0}^{\infty}\frac{dt}{t^{3/2}}\, e^{\frac{-x^{2}}{4t}}e^{-t}\,(Tr'\, e^{\frac{t}{m^{2}}\Delta}),
\end{equation}
where $Tr'$ means the ground state is omitted in the trace
\begin{equation}
  Tr'\, e^{t\Delta}=Tr\, e^{t\Delta}-e^{-\lambda_{0}t}.
\end{equation}
Here $\lambda_{0}$ is the lowest eigenvalue of $-\Delta$. Note that the heat kernel is now given as
\begin{equation}
  Tr'\, e^{t\Delta}\sim \sum_{k=0}^{\infty}a'_{k/2}t^{\frac{k-3}{2}},
\end{equation}
where $a'_{2l+3}=a_{2l+3}-(-1)^{l}\lambda_{0}^{l}/l!$, and $a'_{k}=a_{k}$, otherwise.
The small $x$ asymptotic expansion of $F_{1}$ is still given by (\ref{b}) with the only difference that instead of $\{a_{k}\}$ coefficients we now have $\{a'_{k}\}$ coefficients.
Moreover, since $F_{1}(x)<F(x)$ the estimates (\ref{ineq1}) and (\ref{largeF}) for $F(x)$ are also valid for $F_{1}(x)$.

The high temperature expansion can now be obtained by starting with the expression
\begin{equation}
 Q_{e}=\frac{1}{\sqrt{\pi}} \sum\limits_{k=1}^\infty e^{-\alpha k x} f_{\alpha}(kx),
\end{equation}
where
\begin{equation}
  f_{\alpha}(x)= e^{\alpha  x} x \sinh(\xi x) F(x),
\end{equation}
and proceeding as in the case of free energy. The small $x$ asymptotic of $f_{\alpha}$ is obtained by combining the asymptotic of $F(x)$ with the Taylor expansions of $\sinh(\xi x)$  and  $e^{\alpha  x}$:
\begin{equation}
 f_{\alpha}(x)\sim c_{-2}x^{-2}+c_{-1}x^{-1}+c_{0}+\ldots
\end{equation}
where the first few $c$ coefficients are given in terms of $b$ coefficients as
\begin{eqnarray}
  c_{-2} &=& \xi b_{-4} \\
  c_{-1} &=& \xi b_{-3}+\alpha\xi b_{-4} \\
  c_{0} &=& \xi b_{-2}+\alpha \xi b_{-3}+\left(\frac{\xi\alpha^{2}}{2}+\frac{\xi^{3}}{6}\right)b_{-4}.
 \end{eqnarray}
On the other hand using (\ref{largeF}), for $x\rightarrow \infty$ we get
\begin{equation}\label{large2}
  f_{\alpha}(x) =O\left(x^{-1/2}\,e^{-(1-\xi-\alpha)x}\right).
\end{equation}
Now the Mellin transform of $Q_{e}$ is given by
\begin{equation}
  (\mathcal{M}Q_{e})(s)=\frac{1}{\sqrt{\pi}}\zeta(s)(\mathcal{LM}f_{\alpha})(s,\alpha).
\end{equation}
\begin{eqnarray}
  (\mathcal{LM}f_{\alpha})(s,\alpha)&=&\int_{0}^{\infty}dx\,x^{s-1}\,e^{-\alpha x}\left[ f(x)-c_{-2}x^{-2}\right]+\int_{0}^{\infty}dx\,x^{s-1}\,e^{-\alpha x}c_{-2}x^{-2}\nonumber\\
    &=&\int_{0}^{\infty}dx\,x^{s-1}\,e^{-\alpha x}\left[ f(x)-c_{-2}x^{-2}\right]+c_{-2}\alpha^{2-s}\Gamma(s-2)\nonumber\\&=&
    \ldots.
\end{eqnarray}
Proceeding as in the calculation of the free energy we get the expansion
\begin{equation}
  (\mathcal{LM}f_{\alpha})(s,\alpha)\sim c_{-2}\alpha^{2-s}\Gamma(s-2)+c_{-1}\alpha^{1-s}\Gamma(s-1)+c_{0}\alpha^{-s}\Gamma(s)+\ldots
\end{equation}
Note that all the poles are simple. The pole at $s=2$ comes from the $c_{-2}$ term, while the one at $s=1$ comes from both $c_{-2}$ and $c=-1$ terms, etc.

For $s=2$:
\begin{equation}
  (\mathcal{M}Q_{e})(s)\asymp \frac{1}{\sqrt{\pi}}\zeta(2) (\xi b_{-4})\,\frac{1}{s-2}
\end{equation}
Corresponding asymptotic is
\begin{eqnarray}
  Q_{e}^{(2)}= \pi^{-1/2} \zeta(2)(\xi b_{-4})(\beta m)^{-2}
\end{eqnarray}

For $s=1$:
\begin{eqnarray}
   (\mathcal{M}Q_{e})(s) &\sim &\pi^{-1/2} \left(\frac{1}{s-1}+\gamma\right)\left(FP(\mathcal{LM}f)(s)\left.\right|_{s\rightarrow 1}+\frac{\xi b_{-3}}{s-1}\right)\nonumber  \\
   &\asymp &\pi^{-1/2}\left[ \frac{\xi b_{-3}}{(s-1)^{2}}+[\gamma \xi b_{-3} +FP(\mathcal{LM}f)(s=1,\alpha)]\frac{1}{s-1}\right].\nonumber\\
\end{eqnarray}

Finally for $s=0$:
\begin{equation}
(\mathcal{M}Q_{e})(s)\asymp \frac{1}{\sqrt{\pi}}\zeta(0)\left[\frac{\xi^{3}}{6}b_{-4}+\xi b_{-2} \right]\,\frac{1}{s}.
\end{equation}

Thus we arrive at the high temperature expansion
\begin{equation}
  Q_{e}\sim Q_{e}^{(2)}+Q_{e}^{(1)}+\ldots
\end{equation}
with
\begin{eqnarray}
  Q_{e}^{(2)}= \pi^{-1/2} \zeta(2) \xi b_{-4}(\beta m)^{-2},
\end{eqnarray}
\begin{eqnarray}
  Q_{e}^{(1)} &=& -\pi^{-1/2}(\xi b_{-3})(\beta m)^{-1}\log (\beta m)^{-1}+\nonumber\\
  &&+\pi^{-1/2}\left[\gamma (\xi b_{-3})+FP(\mathcal{LM}f_{\alpha})(1,\alpha)\right](\beta m)^{-1},
 \end{eqnarray}
Using (\ref{b}) we can express the above in terms of the heat kernel coefficients as
\begin{eqnarray}\label{ncharged3}
 Q_{e}^{(2)} &=& \frac{16}{\sqrt{\pi}}\zeta(2)  \mu m^{2} a_{0}  (\beta m)^{-2} \nonumber\\
 Q_{e}^{(1)} &=& - \left( 4 m\mu a_{1/2} \right)  (\beta m)^{-1} \log (\beta m)^{-1} +\nonumber\\
   &&+\left[ \gamma 4 m \mu a_{1/2}   +    \frac{FP(\mathcal{LM}f_{\alpha})(1,\alpha)}{\sqrt{\pi}} \right] (\beta m)^{-1}.
\end{eqnarray}

In higher dimensions, i. e. for  $d>3$,  a similar analysis gives
\begin{eqnarray}\label{nchargedd}
Q_{e} &\sim & \zeta(d-1) \,\frac{(2m)^{d-1}}{\sqrt{\pi}}  \,4\mu \,\Gamma\left(\frac{d+1}{2} \right)\,  a_{0} \,(\beta m)^{-(d-1)} \nonumber\\
       &+& \zeta(d-2)\, \frac{(2m)^{d-2}}{\sqrt{\pi}}  \,4\mu \, \Gamma\left(\frac{d}{2} \right)\,a_{1/2} \,(\beta m) ^{-(d-2)}\nonumber\\
       &+& \zeta(d-3)\,\frac{(2m)^{d-3}}{\sqrt{\pi}}\,4\mu\,\Gamma\left(\frac{d-1}{2} \right) \left[ \left( -m^{2}+\frac{(d-1)}{3}\mu^{2}\right)a_{0} +  a_{1}\right]   \,(\beta m)^{-(d-3)}\nonumber\\
\end{eqnarray}

To investigate the effect of the boundary on the critical temperature we start by determining $T$ as a function of $N_{0}$, the number of particles in the ground state, for a fixed number of particles $N$. Let us first consider the case $d=3$. Taking into account the leading boundary contribution in (\ref{ncharged3}) we arrive at the equation
\begin{equation}\label{eq1}
  \frac{2V}{\pi^{2}}\zeta(2)\mu m^{2}(\beta m)^{-2}- b\frac{A}{4\pi}\mu m(\beta m)^{-1}\log (\beta m)^{-1}=N-N_{0}.
\end{equation}
In the case of condensation from
\begin{equation}
  Q_{0}=\frac{1}{e^{\beta(m-\mu)}-1}-\frac{1}{e^{\beta(m+\mu)}-1}
\end{equation}
we get
\begin{equation}\label{mu}
  \mu=\pm m-\frac{1}{\beta q_{0}V}+O(V^{-2}).
\end{equation}
Here and below $q_{0}=Q_{0}/V$ and $q=Q/V$. If we set $y=k_{B}T/m$ and divide both sides of (\ref{eq1}) by $V$
\begin{equation}
  \frac{2}{\pi^{2}}\zeta(2)\mu m^{2}y^{2}- \frac{r\varepsilon}{4\pi}\mu m y\log y=q-q_{0}.
\end{equation}
Here $\varepsilon=A/V$ is the small parameter in terms of which we can expand the solution perturbatively
\begin{equation}\label{ypert}
  \frac{k_{B}T}{m}=y=y_{0}+\varepsilon y_{1}+\ldots.
\end{equation}
Now we see that the sign of $q-q_{0}$ is controlled by the sign of $\mu$; $q>q_{0}$ for $\mu>0$ and $q<q_{0}$ for $\mu<0$.
The two cases can be worked out simultaneously. Since the error in replacing $\mu$ by $\pm m$ is $O(V^{-1})$ we have (here we are assuming that the geometry of our box $B$ is regular in the sense that $1/V << A/V$)
\begin{equation}
  \frac{2}{\pi^{2}}\zeta(2) m^{3}y^{2}-r \frac{\varepsilon}{4\pi} m^{2} y\log y=|q-q_{0}|.
\end{equation}
Then the solution is given as
\begin{eqnarray}
  y_{0}&=&\left( \frac{\pi^{2}(|q-q_{0}|)}{2\zeta(2)m^{3}}\right) ^{1/2}\\
  y_{1}&=&\frac{r\pi}{32 m \zeta(2)}\log \frac{\pi^{2}(|q-q_{0}|)}{2\zeta(2)m^{3}}.
\end{eqnarray}
At this point it is tempting to set $n_{0}=0$ to determine the critical temperature, however this is an inconsistent step for finite $V$, since it implies a divergent $\mu$, and is justified only in the thermodynamic limit i.e. after taking the $V\rightarrow \infty$ limit. This situation is similar to the non-relativistic case discussed in \cite{T4}.

The analysis of the higher dimensional case is similar. Setting $y=(\beta m)^{-1}$, $\epsilon=A/V$, $\mu=m$ and making the perturbative expansion (\ref{ypert}) as in the three dimensional case, we get the results;
\begin{eqnarray}
y_{0}=\left[ \frac{(|q-q_{0}|)\,(\sqrt{\pi})^{d+1}}{m\,2^{d}\,\zeta(d-1)\,\Gamma \left( \frac{d+1}{2}\right) }\right] ^{\frac{1}{d-1}}
\end{eqnarray}
\begin{eqnarray}
y_{1}=\pm \frac{\sqrt{\pi}}{4m(d-1)}\,\frac{\zeta(d-2)}{\zeta(d-1)}\,\frac{\Gamma \left( \frac{d}{2}\right) }{\Gamma \left( \frac{d+1}{2}\right) }
\end{eqnarray}
for $d$ dimensional charged boson.

\section{Free Energy of Neutral Bosons}

Starting, as in the charged case, with the subordination identity (\ref{sub}) we can write the free energy for a gas of neutral bosons
\begin{equation}
A=\frac{1}{\beta} \sum \limits_{\sigma} \left[ \log \left( 1- e^{-(\beta^{2} \lambda_{\sigma}+\beta^{2}m^{2})^{1/2} + \beta\mu} \right) \right]
\end{equation}
as
\begin{eqnarray}
  A &=& -\sum\limits_{k=1}^\infty \frac{m}{2\sqrt{\pi}}e^{k\beta \xi m}F(k\beta m)\nonumber\\
  &=& \frac{-m}{2\sqrt{\pi}}\sum\limits_{k=1}^\infty e^{-\alpha k x}g(kx),
\end{eqnarray}
where
\begin{equation}
  g_{\alpha}(x)=e^{(\alpha + \xi) x} F(x)\sim d_{-4}x^{-4}+ d_{-3}x^{-3}+\ldots \;\;\;\;x\rightarrow 0.
\end{equation}
In this case $d$ coefficients in terms of $b$ coefficients are;
\begin{eqnarray}
d_{-4} &=& b_{-4} \nonumber\\
d_{-3} &=& b_{-3}+b_{-4}(\alpha+\xi) \nonumber\\
d_{-2} &=& b_{-2}+b_{-3}(\alpha+\xi)+b_{-4}\frac{(\alpha+\xi)^{2}}{2} \nonumber\\
d_{-1} &=&b_{-1}+b_{-2}(\alpha+\xi)+b_{-3}\frac{(\alpha+\xi)^{2}}{2}+ b_{-4}\frac{(\alpha+\xi)^{3}}{6} \nonumber\\
d_{0}  &=& b_{-1}(\alpha+\xi)+b_{-2}\frac{(\alpha+\xi)^{2}}{2}+ b_{-3}\frac{(\alpha+\xi)^{3}}{6}+ b_{-4}\frac{(\alpha+\xi)^{4}}{24}
\end{eqnarray}
Now the Mellin transform is
\begin{equation}
  (\mathcal{M}A)(s)= \frac{-m}{2\sqrt{\pi}}\zeta(s)(\mathcal{LM}g_{\alpha})(s,\alpha),
\end{equation}
where
\begin{equation}
  (\mathcal{LM}g)(s,\alpha)=\int_{0}^{\infty}dx\,x^{s-1}e^{-\alpha x}g_{\alpha}(x).
\end{equation}
The singular asymptotic can be worked out exactly as in the charged case. Thus we have
\begin{equation}
  (\mathcal{LM}g)(s,\alpha)\sim d_{-4}(\alpha)^{4-s}\Gamma(s-4)+d_{-3}(\alpha)^{3-s}\Gamma(s-3)+\ldots.
\end{equation}
Now the poles are at $s=4,3,2,1,\ldots$. Again the pole at $s=4$ comes from $d_{-4}$ term, the pole at $s=3$ comes from both $d_{-4}$ and $d_{-3}$ terms, etc. We stress that the equation is in the same form as in  the charged case. Let us express the free energy as;
\begin{equation}
A \sim A^{(4)}+A^{(3)}+A^{(2)}+A^{(1)}+\ldots
\end{equation}
where
\begin{eqnarray}
 A^{(4)}&=& \frac{-m}{2\sqrt{\pi}}\zeta(4)(16 m^{3} a_{0}) (\beta m)^{-4} \nonumber\\
 A^{(3)}&=& \frac{-m}{2\sqrt{\pi}}\zeta(3) (16 \mu m^{2} a_{0} +4\sqrt{\pi}m^{2}a_{1/2}) (\beta m)^{-3} \nonumber\\
 A^{(2)}&=& \frac{-m}{2\sqrt{\pi}}\zeta(2) \left[ 4(2m \mu^{2}-m^{3})a_{0} + 4\sqrt{\pi} \mu m a_{1/2} + 4 m a_{1}\right]   (\beta m)^{-2} \nonumber\\
 A^{(1)}&=&\frac{m}{2\sqrt{\pi}} \left[ \left(\frac{8}{3} \mu^{3}-4 m^{2}\mu \right)  a_{0}+ 2\sqrt{\pi} (\mu^{2}-m^{2})a_{1/2} +4\mu a_{1} +2\sqrt{\pi}\, a_{3/2} \right]\nonumber\\  & \times &(\beta m)^{-1} \log (\beta m)^{-1}  \nonumber\\
 &-& \frac{m}{2\sqrt{\pi}} \left( \gamma \left[ \left(\frac{8}{3} \mu^{3}-4 m^{2}\mu \right)  a_{0}+ 2\sqrt{\pi} (\mu^{2}-m^{2})a_{1/2} +4\mu a_{1} +2\sqrt{\pi}\, a_{3/2} \right]   \right. \nonumber\\
  &+& \left. FP(\mathcal{LM}g_{\alpha})(1,\alpha)  \right) (\beta m)^{-1}.
\end{eqnarray}

Similarly, for higher dimensions, i.e. for  $d>3$, we get
\begin{eqnarray}
A &\sim &  -\zeta(d+1) \,\frac{(2m)^{d+1}}{2\sqrt{\pi}}  \, \Gamma\left(\frac{d+1}{2} \right) \,a_{0}\,(\beta m)^{-(d+1)} \nonumber \\
&-&\zeta(d) \,\frac{(2m)^{d}}{2\sqrt{\pi}} \left[ 2\mu\,\Gamma\left(\frac{d+1}{2} \right)\,a_{0}+\Gamma\left(\frac{d}{2} \right)\,a_{1/2}\right] \,(\beta m)^{-d} \nonumber\\
&-& \zeta(d-1) \,\frac{(2m)^{d-1}}{2\sqrt{\pi}}\,\Gamma\left(\frac{d-1}{2}\right)  \left[ \left(-m^{2}+(d-1)\mu^{2}\, \right) \, a_{0}  \right.\nonumber\\
  &-& \left. 2\mu\, \Gamma\left(\frac{d}{2} \right)\,a_{1/2} +  a_{1}  \right](\beta m)^{-(d-1)}
\end{eqnarray}

\section{Number of Neutral Bosons}
With the subordination identity (\ref{sub}) we get for a gas of neutral Bosons
\begin{eqnarray}
N_{e}&=&\sum\limits_{\sigma\neq 0}\frac{1}{e^{\beta( \lambda_{\sigma}+m^{2})^{1/2}-\beta\mu}-1}\\
  &=& \sum\limits_{k=1}^\infty \frac{(k\beta m)e^{k\beta \xi m}}{2\sqrt{\pi}}F(k\beta m)\nonumber\\
  &=& \frac{1}{2\sqrt{\pi}}\sum\limits_{k=1}^\infty e^{k\xi x}f(kx),
\end{eqnarray}
where
\begin{equation}
  f_{\alpha}(x)=e^{(\alpha+\xi)x}xF(x)\sim d_{-4}x^{-3}+ d_{-3}x^{-2}+\ldots \;\;\;\;x\rightarrow 0.
\end{equation}
Again the Mellin transform is
\begin{equation}
  (\mathcal{M}N_{e})(s)= \frac{1}{2\sqrt{\pi}}\zeta(s)(\mathcal{LM}f)(s,\alpha),
\end{equation}
where
\begin{equation}
  (\mathcal{LM}f_{\alpha})(s,\alpha)=\int_{0}^{\infty}dx\,x^{s-1}e^{\alpha x}f(x).
\end{equation}
The asymptotic can be worked out exactly as in the previous cases with the result
\begin{equation}
  (\mathcal{LM}f_{\alpha})(s,-\xi)\asymp d_{-4}(\alpha)^{3-s}\Gamma(s-3)+d_{-3}(\alpha)^{2-s}\Gamma(s-2)+\ldots.
\end{equation}
Now the poles are at $s=3,2,1,\ldots$. Again the pole at $s=3$ comes from $d_{-4}$ term, the pole at $s=2$ comes from both $d_{-4}$ and $d_{-3}$ terms, etc.
The high temperature expansion is then given by
\begin{equation}
N_{e} \sim N_{e}^{(3)}+N_{e}^{(2)}+N_{e}^{(1)}+\ldots
\end{equation}
where
\begin{eqnarray}
N_{e}^{(3)}&=& \frac{1}{2\sqrt{\pi}}\zeta(3)(16 m^{3} a_{0}) (\beta m)^{-3} \nonumber\\
 N_{e}^{(2)}&=&\frac{1}{2\sqrt{\pi}}\zeta(2) (16 \mu m^{2} a_{0} +4\sqrt{\pi}m^{2}a_{1/2}) (\beta m)^{-2} \nonumber\\
 N_{e}^{(1)}&=&\frac{-1}{2\sqrt{\pi}} \left( 4(2m \mu^{2}-m^{3})a_{0} + 4\sqrt{\pi} \mu m a_{1/2} + 4 m a_{1}\right)  (\beta m)^{-1} \log (\beta m)^{-1}  \nonumber\\
 &+& \frac{1}{2\sqrt{\pi}} \left[\gamma \left(4(2m \mu^{2}-m^{3})a_{0} + 4\sqrt{\pi} \mu m a_{1/2} + 4 m a_{1} \right)  +FP(\mathcal{LM}f_{\alpha})(1,\alpha) \right] (\beta m)^{-1}.\nonumber\\
\end{eqnarray}

In higher dimensions, i.e. for  $d>3$, a similar analysis yields
\begin{eqnarray}
N_{e} &\sim & \zeta(d) \, \frac{(2m)^{d}}{\sqrt{\pi}} \,\Gamma\left(\frac{d+1}{2} \right) \, a_{0} \, (\beta m)^{-d} \nonumber\\
&+& \zeta(d-1)\, \frac{(2m)^{d-1}}{\sqrt{\pi}} \,\left[ 2\mu\,\Gamma\left(\frac{d+1}{2} \right)\,a_{0}+\Gamma\left(\frac{d}{2} \right)\,a_{1/2}\right] \,(\beta m)^{-d+1}  \nonumber\\
&+& \zeta(d-2)\,\frac{(2m)^{d-2}}{\sqrt{\pi}} \Gamma\left(\frac{d-1}{2}\right) \left[ \left(-m^{2} +(d-1)\mu^{2}\right) \, a_{0}+\right.\nonumber\\  &+& 2\mu\, \Gamma\left(\frac{d}{2} \right)\,a_{1/2} \left.+   a_{1}  \right] (\beta m)^{-d+2}.
\end{eqnarray}

As in the charged case we can determine the dependence of temperature on $n_{0}$ ($n=N/V$ and $n_{e}=N_{e}/V$) using this high temperature expansion. Let us start with the case $d=3$. We again set $y=k_{B}T/m$, $\varepsilon=A/V$ and note
that from
\begin{equation}
 N_{0}=\frac{1}{e^{\beta(m-\mu)}-1}
\end{equation}
we get
\begin{equation}
  \mu \sim m-\frac{1}{\beta n_{0} V}.
\end{equation}
Since the difference $\mu-m$ is of lower order than $\varepsilon=A/V$ we replace $\mu$ by $m$ and get
 \begin{equation}
 \frac{1}{\pi^{2}}\zeta(3) m^{3} y^{3}+\zeta(2)\left[\frac{1}{\pi^{2}} m^{3}\pm\frac{\varepsilon}{8\pi}m^{2}\right]y^{2}=n-n_{0}
\end{equation}
The perturbative expansion $y=y_{0}+\varepsilon y_{1}+\ldots $ yields
\begin{eqnarray}
  \zeta(3)y_{0}^{3}+ \zeta(2)y_{0}^{2}=\frac{\pi^{2}(n-n_{0})}{m^{3}},
\end{eqnarray}
and
\begin{equation}
  \frac{ 3\zeta(3) m^{3}}{\pi^{2}}y_{0}^{2}y_{1}+ \frac{ 2\zeta(2) m^{3}}{\pi^{2}}y_{0}y_{1}\pm  \frac{ \zeta(2) m^{2}}{8\pi}y_{0}^{2}=0.
\end{equation}
The last equation can easily be solved for $y_{1}$ in terms of $y_{0}$
\begin{equation}
  y_{1}=\mp \frac{\zeta(2)\pi}{8m}\frac{y_{0}}{3\zeta(3)y_{0}+2\zeta(2)}.
\end{equation}
On the other hand the first equation being a cubic has an exact solution, which however is not very transparent. Instead we can solve it by the method of detailed balance \cite{Miller}, assuming small $m$. Let us write it as
\begin{equation}
  \zeta(3)y_{0}^{3}+\eta \zeta(2)y_{0}^{2}=\frac{\pi^{2}(n-n_{0})}{m^{3}}.
\end{equation}
Here $\eta$ is an expansion parameter which will be set equal to $1$ at the end of the calculation. Now plugging in the perturbation expansion $y_{0}=z_{0}+\eta z_{1}+\ldots$ we get
\begin{equation}
  y_{0}=\frac{((\zeta(3))^{-1}\pi^{2}(n-n_{0}))^{1/3}}{m}-\frac{\zeta(2)}{3\zeta(3)}+\ldots.
\end{equation}
By working few more terms of the perturbation series one can easily see that it is an expansion in powers of $m$ and the neglected terms are in fact $O(m)$. Note that the leading term above is the result obtained in \cite{Begun0}.

The analysis can be generalized to higher dimensions,  with the result
\begin{eqnarray}
y_{0}=\left[ \frac{(n-n_{0})\,(\sqrt{\pi})^{d+1}}{m^{d}\,\zeta(d)\,\Gamma \left( \frac{d+1}{2}\right) }\right] ^{\frac{1}{d}}-\frac{\zeta(d-1)}{d\,\zeta(d)},
\end{eqnarray}
\begin{eqnarray}
y_{1}=\mp  \frac{\sqrt{\pi}}{4m}\,\frac{\Gamma \left( \frac{d}{2}\right) }{\Gamma \left( \frac{d+1}{2}\right) }\,\frac{\zeta(d-1)\, y_{0}}{\left[ d\,\zeta(d)\,y_{0}+(d-1)\,\zeta(d-1)\right] },
\end{eqnarray}
where $d>3$.

\section*{Acknowledgement}

This work is supported by Bo\u{g}azi\c{c}i University BAP Project No. 6942.

\section*{Appendix A: Asymptotic Expansion of $F(x)$}
We will now derive the asymptotic expansion, more precisely only the singular part of the asymptotic expansion, of the function $F(x)$ as $x\rightarrow 0$ and thus determine the coefficients (\ref{b}).

Write
\begin{eqnarray}
  F(x)&=&\int_{0}^{\infty}dt\,t^{-3/2}\,e^{-\frac{x^{2}}{4t}-t}\left[Tr'\, e^{\frac{\Delta}{m^{2}}t}-t^{-3/2}\sum_{n=0}^{4}a_{n/2}(tm^{-2})^{n/2}\right]\nonumber\\
 && +\int_{0}^{\infty}dt\,t^{-3/2}\,e^{-\frac{x^{2}}{4t}-t}(tm^{-2})^{-3/2}\sum_{n=0}^{4}a_{n/2}(tm^{-2})^{n/2}.
\end{eqnarray}
Now split the first integral as
\begin{equation}
 \int_{0}^{1}+\int_{1}^{\infty} dt\,t^{-3/2}\,e^{-\frac{y^{2}}{4t}-t}\left[Tr'\, e^{\frac{\Delta}{m^{2}}t}-t^{-3/2}\sum_{n=0}^{4}a_{n/2}(tm^{-2})^{n/2}\right].
\end{equation}
The first integral is absolutely convergent as $x\rightarrow 0$ since because of the heat kernel expansion the term in square brackets is $O(t^{-1})$ and therefore the integrand is
$O(t^{-1/2})$, as $t\rightarrow 0$. On the other hand in the second integral $e^{-t}$ term and the boundedness of $Tr'\, e^{\frac{\Delta}{m^{2}}t}$ as $t\rightarrow \infty$ imply convergence.

Thus we see that the singular asymptotic of $F(x)$ as $x \rightarrow 0$ is given by
\begin{equation}
  F(x)\asymp \sum_{n=0}^{4}m^{3-n}\,a_{n/2}F_{n}(x),
\end{equation}
where
\begin{equation}
  F_{n}(x)= \int_{0}^{\infty} dt\,t^{n/2-3}\,e^{-\frac{x^{2}}{4t}-t}=2\left(\frac{x}{2}\right)^{-(2-\frac{n}{2})}K_{2-\frac{n}{2}}(x).
\end{equation}
Here $K$'s are the modified Bessel functions. Using the series expansion thereof for non-integer $\nu$ \cite{Abr}
\begin{eqnarray}
  K_{\nu}(x) &=& \frac{\pi \csc (\nu \pi)}{2}\left[\sum_{k=0}^{\infty}\frac{1}{\Gamma(k-\nu+1)k!}\left(\frac{x}{2}\right)^{2k-\nu}-(\nu\rightarrow -\nu)\right],
\end{eqnarray}
and for integer $n$
\begin{eqnarray}
  K_{n}(x)&=&\frac{1}{2}\sum_{k=0}^{n-1}(-1)^{k}\frac{(n-k-1)!}{k!}\left(\frac{x}{2}\right)^{2k-n}+\nonumber\\
  &&+(-1)^{n+1}\sum_{k=0}^{\infty}\frac{\left(\frac{x}{2}\right)^{2k+n}}{k!(n+k)!}\left[\log\frac{x}{2}-\frac{1}{2}\psi(k+1)-\frac{1}{2}\psi(n+k+1)\right],
\nonumber \\
\end{eqnarray}
collecting the relevant terms we get
\begin{eqnarray}
  F(x) &\asymp& m^{3}a_{0}\left(\frac{x}{2}\right)^{-4}+m^{2}\frac{\sqrt{\pi}}{2}a_{1/2}\left(\frac{x}{2}\right)^{-3}+(-m^{3}a_{0}+ma_{1})\left(\frac{x}{2}\right)^{-2}\nonumber \\
   &=&(-m^{2}\sqrt{\pi}a_{1/2}+\sqrt{\pi}a_{3/2})\left(\frac{x}{2}\right)^{-1}+(-m^{3}a_{0}+2ma_{1}-m^{-1}a_{2})\log\frac{x}{2}.\nonumber\\
\end{eqnarray}
From this we get the $b$ coefficients as given in (\ref{b}).

\section*{Appendix B: Asymptotic Expansion Through Mellin Transform}
Here we will discuss the validity of the Mellin transform to generate the high temperature expansion. We will consider only the free energy for the charged bosons. The other cases can be treated similarly. Consider the Mellin transform  $(\mathcal{M}f)(s)$ of a function $f(x)$. In general $(\mathcal{M}f)(s)$ is holomorphic on a strip $a<\Re s<b$. It is assumed that $(\mathcal{M}f)(s)$ has a meromorphic extension into $\Re s\leq a$ with poles at $a=w_{1}\geq w_{2} \geq \ldots $. Let $A(w_{r},k_{r})$ be the residue with multiplicity $k_{r}$ of the pole at $w_{r}$.

The inverse Mellin transform is given by
\begin{equation}
  f(x)=\int_{c-i\infty}^{c+i\infty}ds\,x^{-s}\,(\mathcal{M}f)(s),
\end{equation}
where $a<c<b$. The asymptotic expansion of $f(x)$ is obtained by closing the above contour into a rectangular one where the added vertical contour at $\Re s=\sigma_{0}<c$ lies between two consecutive poles, say $w_{r}$ and $w_{r+1}$ (assuming the real parts of these poles are distinct).The expansion is derived by completing the contour in the inverse Mellin transform into a rectangular contour enclosing the poles of the meromorphic extension of  $(\mathcal{M}f)(s)$. For this to be an asymptotic expansion the contributions of the edges used to close the contour must be negligible. The following sufficient condition is suitable for our purposes. Suppose $\sigma$ is a real number lying between two consecutive poles of $(\mathcal{M}f)(s)$ and let us suppose
\begin{equation}
 \int_{-\infty}^{\infty}d\tau\,|(\mathcal{M}f)(\sigma+i\tau)|<\infty.
\end{equation}
Then what we calculate is  an asymptotic expansion of the function $f(x)$. To summarize,
if
\begin{equation}\label{as1}
  \lim_{|\tau|\rightarrow\infty}|(\mathcal{M}f)(\sigma+i \tau)|= 0 \;\;\;\textrm{for}\;\;\;\sigma_{0}\leq\sigma \leq c,
\end{equation}
 and
\begin{equation}\label{as2}
  \int_{-\infty}^{\infty}d\tau\,|(\mathcal{M}f)(\sigma_{0}+i \tau)|<\infty,
\end{equation}
then we find
\begin{equation}
  f(x)\sim \sum_{i=1}^{r}\sum_{k}^{k_{r}}A(w_{r},k_{r})x^{-w_{r}}(\log x)^{k_{r}}.
\end{equation}

We will apply this to our case where $f=A$, the free energy.
\begin{equation}
  (\mathcal{M}A)(s)=\frac{-m}{\sqrt{\pi}}\zeta(s)(\mathcal{LM}g_{\alpha})(s,\alpha).
\end{equation}
Now $|\zeta(\sigma+i\tau)|$ increases algebraically as $|\tau|\rightarrow \infty$ \cite{Whit} (provided $a$ is not close to $0$, which is true in our case). If we can show that for large $|\tau|$, $|(\mathcal{LM}g_{\alpha})(s,\alpha)|=O(P(|\tau|)e^{-q|\tau|})$ where $P$ is a polynomial and $q>0$, then the hypothesis of the theorem will be satisfied.

 The meromorphic extension of $(\mathcal{LM}g_{\alpha})$ we are interested in is
\begin{equation}
    \int_{0}^{\infty}dx\,x^{s-1}e^{-\alpha x}\left[g_{\alpha}(x)-\sum_{j=-4}^{-1} c_{j}x^{j}\right]+\sum_{j=-4}^{-1} c_{j}\alpha^{-j-s}\Gamma (s+j).
\end{equation}
Here $s=\sigma+i\tau=|s|e^{i\phi}$, $\sigma=1/2$. For later convenience we also define $\hat{s}=s/|s|=p+iq$.
From \cite{Abr}
\begin{equation}
    |\Gamma(\sigma+i\tau)|=O(|\tau|^{\sigma-\frac{1}{2}}e^{-\frac{1}{2}\pi|\tau|})
\end{equation}
we see that the last four terms have the the desired property of fast decay. So let us concentrate on the first term.

Rotate our contour $(0,\infty)$ by $\phi<\pi/2$. Along this new contour $C$ the integral is equal to
\begin{equation}
  \int_{C}dz\ldots=\hat{s}^{s}\int_{0}^{\infty}dx x^{\sigma-1} e^{-\alpha\hat{s} x}[g(\hat{s}x)-\sum_{j=-4}^{-1} c_{j}x^{j}].
\end{equation}
For real $x$ let us write (\ref{F}) as
\begin{equation}
    F(x)=\frac{1}{x^{1/2}}\int_{0}^{ \infty}\frac{du}{u^{3/2}}\,e^{-\frac{x}{4u}}e^{-xu}Tr e^{\frac{x\Delta}{m^{2}}u}.
\end{equation}
Using this expression to continue $F$ analytically to $\Re z>0$ we get the following analytic continuation of $g$
\begin{equation}
    g_{\alpha}(z)=e^{\alpha z} \cosh (\xi z)F(z).
\end{equation}
We know that for real $x$, $ g_{\alpha}(x)-\sum_{j=-4}^{-1} c_{j}x^{j}=O(1)$. Assuming the heat kernel expansion can be continued to complex time and repeating the derivation of Appendix A with $x$ replaced by $\hat{s} x$ we get
\begin{equation}
    g(\hat{s}x)-\sum_{j=-4}^{-1} c_{j}x^{j}=O(1).
\end{equation}
Thus
\begin{equation}
    \left| \int_{C}dz\ldots\right|\leq C |\hat{s}^{s}|\frac{\Gamma(\sigma)}{(\alpha p)^{\sigma}}.
\end{equation}
Now
\begin{equation}
  |\hat{s}^{s}|=\frac{|s^{s}|}{|s|^{\sigma}}=e^{-\tau \phi}=e^{-|\tau||\phi|}.
\end{equation}
Here we noted $\tau\phi=|\tau||\phi|$. So we have the desired fast decay property.

We must also show that the contribution to the result of the big circular arc $C_{R}$ of radius $R$ that connects the original contour $(0,\infty)$ to $C$ is negligibly small.
Setting $z=R\,e^{i\theta}$ we get
\begin{equation}
    |g_{\alpha}(z)|\leq e^{\alpha R\cos\theta}\cosh (\xi R\cos\theta)|F(z)|.
\end{equation}
Now notice that
\begin{equation}
|F(z)|\leq R^{-1/2}F(R\cos\theta).
\end{equation}
Thus we obtain the bound
 \begin{equation}
    |g_{\alpha}(z)|\leq R^{-1/2}\,e^{\alpha R\cos\theta}\cosh (\xi R\cos\theta)F(R\cos\theta).
\end{equation}
Using (\ref{largeF}) in this we get
\begin{equation}
    |g_{\alpha}(z)|\leq c R^{-1/2}\,e^{\alpha R\cos\theta}\cosh (\xi R\cos\theta)(R\cos\theta)^{-3/2}e^{-R\cos\theta}
\end{equation}
Since $-1+\alpha+\xi<0$ and $-1+\alpha-\xi<0$ the product of exponential terms is less than $1$. Thus we arrive at
\begin{equation}
   |g_{\alpha}(z)|\leq c(R\cos\theta)^{-3/2}.
\end{equation}

Now along $C_{R}$
\begin{eqnarray}
\left|\int_{C_{R}}dz\ldots\right| &\leq & R^{\sigma}\int_{0}^{\phi}d\theta e^{-\alpha R \cos\theta}\left[|g_{\alpha}(z)|+\sum_{j=-4}^{-1} |c_{j}|R^{j}\right]\nonumber\\
&\leq &R^{\sigma}e^{-R\cos\phi}\int_{0}^{\phi}d\theta \left[c(R\cos\theta)^{-3/2}+\sum_{j=-4}^{-1} |c_{j}|R^{j}\right].
\end{eqnarray}
This shows that the integral along $C_{R}$ vanishes as $R\rightarrow\infty$.


\end{document}